\input{aipcheck}

\documentclass[
    ,final            
  ]
  {aipproc}

\layoutstyle{8x11double}
\usepackage{color}

\begin{document}

\title{Scaling of the normal coefficient of restitution for wet impacts}


\author{Thomas M\"uller}{
  address={Experimentalphysik V, Universit\"at Bayreuth, 95440 Bayreuth, Germany}
}
\author{Frank Gollwitzer}{
  address={Experimentalphysik V, Universit\"at Bayreuth, 95440 Bayreuth, Germany}
}
\author{Christof A. Kr\"ulle}{
  address={Maschinenbau und Mechatronik, Hochschule Karlsruhe - Technik und Wirtschaft, 76133 Karlsruhe, Germany}
}
\author{Ingo Rehberg}{
  address={Experimentalphysik V, Universit\"at Bayreuth, 95440 Bayreuth, Germany}
} 
\author{Kai Huang}{
  address={Experimentalphysik V, Universit\"at Bayreuth, 95440 Bayreuth, Germany}
}

\begin{abstract}

A thorough understanding of the energy dissipation in the dynamics of wet granular matter is essential for a continuum description of natural phenomena such as debris flow, and the development of various industrial applications such as the granulation process. The coefficient of restitution (COR), defined as the ratio between the relative rebound and impact velocities of a binary impact, is frequently used to characterize the amount of energy dissipation associated. We measure the COR by tracing a freely falling sphere bouncing on a wet surface with the liquid film thickness monitored optically. For fixed ratio between the film thickness and the particle size, the dependence of the COR on the impact velocity and various properties of the liquid film can be characterized with the Stokes number, defined as the ratio between the inertia of the particle and the viscosity of the liquid. Moreover, the COR for infinitely large impact velocities derived from the scaling can be analyzed by a model considering the energy dissipation from the inertia of the liquid film.
\end{abstract}

\maketitle



Understanding the energy dissipation associated with particle-particle interactions is crucial for describing the collective behavior of granular matter \cite{Duran00}, i.e., large agglomerations of macroscopic particles. The coefficient of restitution (COR), firstly introduced by Newton as the ratio between relative rebound and impact velocities \cite{Newton1687}, can be used to characterize the energy dissipation at the particle level. This number provides one of the basic ingredients of computer assisted modeling, such as molecular dynamics (MD) simulation, which has been developed into a powerful tool to describe the large scale collective behavior of granular matter in the past decades \cite{Bizon98, bril04}. Besides the energy dissipation from particle-particle interactions, the dissipation arising from the interstitial air or liquid has to be considered when coping with natural phenomena such as dune migration \cite{Bagnold41} or debris flow \cite{Iverson97}, as well as with various industrial 
applications such as granulation process \cite{Iveson01, Antonyuk09}.

The experience of building sand sculptures tells us that the rigidity of a granular material increases as a small amount of a wetting liquid is added. This is largely due to the cohesion arising from the formation of capillary bridges between adjacent particles \cite{Scheel08}. The so-called wet granular matter behaves dramatically different from noncohesive dry granular matter while agitated, with emerging critical behavior, such as phase transitions \cite{Fingerle08} and pattern formations \cite{Huang11}, being traceable to the energy or force scale of a single capillary bridge. In order to gain insights into the dynamical behavior of wet granular matter, it is essential to explore the COR and the associated energy dissipation of wet impacts. A recent investigation reveals that the dependence of the COR on various particle and liquid properties can be scaled with two dimensionless numbers: the ratio between the inertia of the particle and the viscosity of the liquid (Stokes number), and that between the liquid film thickness and the size of the particle \cite{Gollwitzer12}. Here, further experimental results with a different density ratio between the particle and the wetting liquid are presented, in order to test the scaling of the COR with these three dimensionless numbers.


\begin{figure}\label{fig:001}
 \includegraphics[height=.25\textheight]{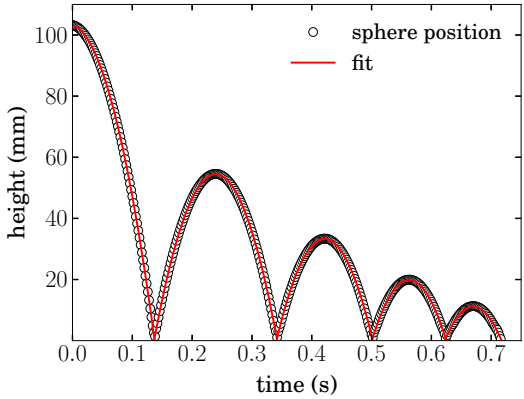}
 \caption{Typical trajectory of a bouncing sphere. The diagram shows a PE sphere of diameter $D=7.9\,$mm bouncing on a glass plate covered with a $190\,\mu$m thick silicone oil (M5) film.}
\end{figure}

We measure the COR by recording a freely falling sphere with a highspeed camera (Photron, Fastcam Super 10K) at a frame rate up to 500\,fps. Using a Hough transformation, the center of the sphere can be tracked in the images and the position of the sphere can be plotted against time (see Fig. \ref{fig:001}). Fitting parabolic curves on the trajectories results in crossing points, which represent the time when the sphere hits the ground. The height of the crossing points varies slightly, owing to the various distance of the bouncing point to the camera. The slopes of the two fitted parabolae at each crossing point yield the impact velocity $v_\mathrm{imp}$ and the rebound velocity $v_\mathrm{reb}$ of the sphere, respectively. Hence the normal coefficient of restitution $e_{\mathrm{n}}=v_\mathrm{reb}/v_\mathrm{imp}$ can be determined for every impact. Polyethylene (PE) spheres with various diameter $D$ and a density $\rho_{\rm p}=0.94\,{\rm g/cm^3}$ are cleaned and pre-wetted before use. Two types of silicone oil (M5 and M50 from Carl Roth) with different dynamic viscosities $\eta=4.6\,{\rm mPa\cdot s}$ for M5 and $48\,{\rm mPa\cdot s}$ for M50, and similar densities $\rho_{\rm liq}=0.93\,{\rm g/cm^3}$ and $0.97\,{\rm g/cm^3}$ for M5 and M50 correspondingly are used. The bottom of the glass container, which has an inner area of $A=100\,{\rm cm^2}$ is leveled within $0.03$\,degrees to ensure a flat surface and a uniform initial liquid film thickness. A sketch of the set-up and a more detailed description can be found in Ref.\,\cite{Gollwitzer12}.

\begin{figure}\label{fig:002}
 \includegraphics[height=.3\textheight]{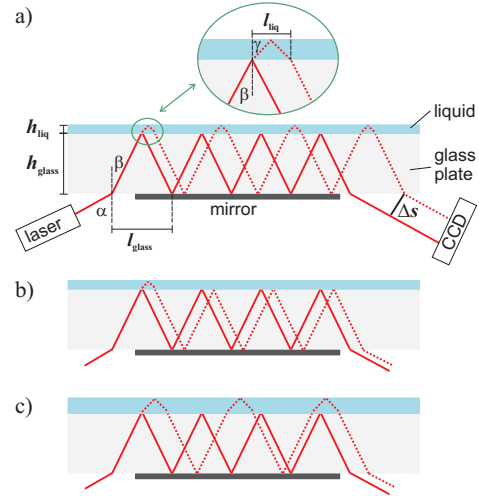}\\
 \caption{a) A sketch of the film thickness determination part of the setup. The solid lines show the beam for a dry container, the dashed lines illustrate the deviation with a liquid layer on top of the glass plate. Number of shifts by passing through the glass plate $j_\mathrm{glass}=4$, number of reflections in the liquid film $j_\mathrm{liq}=4$. b) $j_\mathrm{glass}=4$, $j_\mathrm{liq}=1$. c) $j_\mathrm{glass}=3$, $j_\mathrm{liq}=3$.}
\end{figure}

To investigate the influence of the liquid film on the COR, an accurate determination of its thickness is essential. Here, the thickness is measured optically by detecting the deflection of a laser beam (see Fig. \ref{fig:002} a) guided through the liquid film. The laser beam hits the glass plate with an incident angle of $\alpha$ and a refractive angle $\beta=\arcsin(n_\mathrm{air}\sin\alpha/n_\mathrm{glass})$, where $n_\mathrm{air}$ and $n_\mathrm{glass}$ are the refractive indices of air and glass, respectively. Without a liquid film, a part of the beam is reflected at the glass--air interface and then reflected at a mirror at the bottom of the glass plate. After a few reflections at the glass--air interface (here the number of reflections $j_0=4$, determined by the length of the mirror), the beam passes through the bottom of the glass plate again and is detected by a CCD camera.

With a liquid film of thickness $h_\mathrm{liq}$ on top of the glass plate, one reflection of the beam within the liquid film leads to a horizontal shift of the beam by a distance 
\begin{equation}\label{eq:001}
 l_{\rm liq}=2h_\mathrm{liq}\tan\gamma,
\end{equation}
with the refractive angle in the liquid $\gamma=\arcsin(n_\mathrm{glass}\sin\beta/n_\mathrm{liq})=\arcsin(n_\mathrm{air}\sin\alpha/n_\mathrm{liq})$.

For the typical case that the number of reflections at the air--liquid interface $j_{\rm liq}$ is the same as that in the dry glass plate (Fig. \ref{fig:002} a), the total shift of the laser beam along the horizontal plane is $j_\mathrm{liq}\cdot l_{\rm liq}$ with $j_\mathrm{liq}=j_0$. It is also possible, as shown in Fig. \ref{fig:002} b, to detect reflected beams with a smaller number of passes through the liquid film ($j_\mathrm{liq}<j_0$), due to partial reflections at the glass--liquid interface. If the filling level is too high and consequently the shift of the beam is too big, an additional reduction of reflections in the glass plate to $j_\mathrm{glass}<j_0$ is also possible (Fig. \ref{fig:002} c). In such a case, a negative shift of $(j_0-j_\mathrm{glass})\cdot l_{\rm glass}$ has to be considered, where

\begin{equation}\label{eq:002}
 l_{\rm glass}=2h_\mathrm{glass}\tan\beta
\end{equation}
is a constant value. 

Taking all the above possibilities into account, the reflected beams leave the glass with possible horizontal shifts of $j_\mathrm{liq}\cdot l_{\rm liq}-(j_0-j_\mathrm{glass})\cdot l_{\rm glass}$, which results in
\begin{equation}\label{eq:003}
 \Delta s=\left[j_\mathrm{liq}\cdot l_{\rm liq}-(j_0-j_\mathrm{glass})\cdot l_{\rm glass}\right]\cos\alpha
\end{equation} 
at the CCD. Inserting Eq.~{\ref{eq:001}} and Eq.~{\ref{eq:002}} into Eq.~{\ref{eq:003}}, the liquid film thickness can then be calculated by
\begin{equation}\label{eq:004}
 h_\mathrm{liq}=\frac{\frac{\Delta s}{\cos\alpha}+2(j_0-j_\mathrm{glass})h_\mathrm{glass}\tan\beta}{2j_\mathrm{liq}\tan\gamma}
\end{equation}
Note that for thin films (when $j_0-j_\mathrm{glass}=0$), $h_\mathrm{liq}$ is independent on the properties of the glass plate.

\begin{figure}
 \label{fig:cali1}
 \includegraphics[height=.25\textheight]{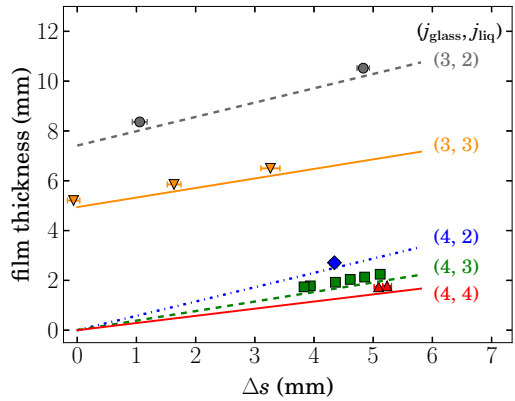}
 \caption{The relation between the water film thickness and the shift of the laser beam $\Delta s$ at the CCD predicted from Eq.~\ref{eq:004} (solid lines) and estimated with $V/A$, the toal liquid volum over the area of the container (data points).}
\end{figure}

Experimentally, the camera captures a series of spots with each of them corresponding to a certain combination of $j_\mathrm{liq}$ and $j_\mathrm{glass}$. The brightest spot corresponds to the case $j_\mathrm{liq}=j_\mathrm{glass}=j_0$, therefore it is commonly used for the film thickness measurement.  Fig.~\ref{fig:cali1} shows a comparison of the film thickness obtained from $V/A$, with $V$ the volume of liquid added, and $h_{\rm liq}$ given by Eq.~\ref{eq:004}. With a certain $V$, the multiple spots detected give rise to multiple $\Delta s$. As $V$ increases, the shift of each spot $\Delta s$ follows a linear growth with various slopes and offsets, which follows the prediction based on Eq.~\ref{eq:004} with various combination of $j_\mathrm{liq}$ and $j_\mathrm{glass}$. The systematic overestimation of the film thickness from $V/A$ for all data sets is presumably due to the volume of the liquid captured in the meniscus of the liquid film $V_{\rm men}$, since the systematic deviation from the estimation of Eq.~\ref{eq:004} does not depend on the parameters $j_\mathrm{liq}$ and $j_\mathrm{glass}$.

Therefore, using the variation of any spot detected by the camera, not only the brightest one, we can measure the liquid film thickness $h_{\rm liq}$. This additional information from the spots with smaller intensity gives the opportunity to expand the range of $h_{\rm liq}$ that can be detected, provided that the free parameters $j_\mathrm{liq}$ and $j_\mathrm{glass}$ have been correctly determined from the first few data points.

\begin{figure}
 \label{cali2}
 \includegraphics[height=.25\textheight]{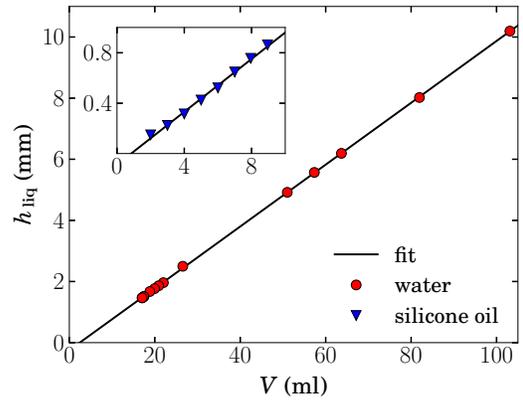}
 \caption{The film thickness obtained with $\Delta s$ and Eq.~\ref{eq:004} as a function of the filling volume of water and silicone oil (inset). The error bars are within the size of the symbols. Linear fits of the data with $h_{\rm liq}=k(V-V_{\rm men})$ yield the same $k=0.01$\,cm$^{-2}$ for both liquids, and $V_{\rm men}=(2.45\pm0.10)$\,ml for water and $(0.84\pm0.15)$\,ml for silicone oil.}
\end{figure}

Figure~\ref{cali2} shows that the optically obtained film thickness grows linearly with the liquid volume $V$, giving rise to a slope $k=1/A$ and an estimation of the meniscus volume $V_{\rm men}$. Therefore, an estimation of $h_{\rm liq}$ with $V/A$ is also appropriate, provided that the meniscus volume is substracted. Nevertheless, the optical way provides a real-time monitoring of the film thickness. This will be helpful in future analysis on surface waves generated by the impact, and in monitoring the loss of liquid due to evaporation. The small deviation from the linear fit for the silicone oil film thickness data (inset of Fig.~\ref{cali2}) indicates a correlation between $V_{\rm men}$ and $h_{\rm liq}$ for thin films, which we leave to future investigations. Figure~\ref{cali2} also shows the possibility to measure the film thickness up to $1$~cm combining the information from various spots detected. The error of $h_{\rm liq}$, on the order of $\approx 10\,\mu m$, arises mainly from the fluctuations of the spot intensity. At a larger film thickness, the spots will become too weak to be detected accurately by the camera. For thin films, silicone oil instead of water is preferable, because its low surface tension prevents the dewetting instability. Thus it is used for the following experiments.


\begin{figure} 
 \label{fig:e_vi}
 \includegraphics[height=.26\textheight]{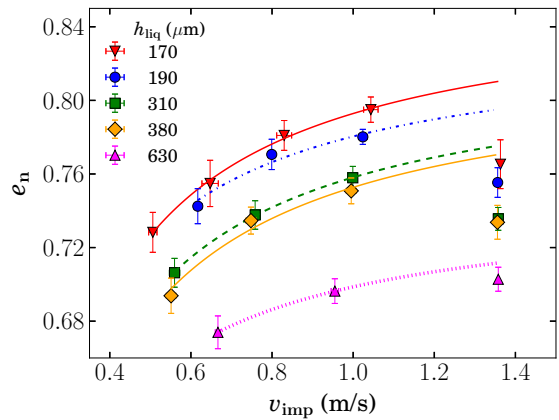}
 \caption{Normal coefficient of restitution $e_{\rm n}$ against the impact velocity $v_{\rm imp}$ for a PE sphere of fixed diameter $D=7.9\,$mm at different silicone oil M5 film thickness $h_{\rm liq}$.}
\end{figure}

Towards a comprehensive understanding of the dependence of the COR for wet impacts on various liquid and particle properties, former experiments with glass beads have revealed that the Stokes number and the dimensionless film thickness $\tilde{h}=h_{\rm liq}/D$ can characterize the influence from inertia and viscous damping on the COR \cite{Gollwitzer12}. The Stokes number is defined as ${\rm St}=\rho_{\rm p}Dv_{\rm imp}/9\eta$. Here, we test the scaling with PE particles, which correspond to a density ratio $\tilde{\rho}=\rho_{\rm liq}/\rho_{\rm p}\approx 1$. Without the liquid, the COR for the PE particle on the glass plate, depending weakly on $v_{\rm imp}$ for the range of interest here, is measured to be $0.88\pm 0.03$, averaged over 368 bouncing events. 

Figure~\ref{fig:e_vi} shows the COR measured with an initial falling height of $9.5$\,cm and various film thicknesses $h_{\rm liq}$. The number of data points for each $h_{\rm liq}$ corresponds to the number of bouncings used for extrapolating the COR. The error bar arises from the statistics over 8-10 runs of the experiment. Except for the first bouncing data (the one with largest $v_{\rm imp}$), the monotonic growth of the COR with $v_{\rm imp}$ could be fitted with $e_{\rm n}=e_{\rm inf}(1-v_{\rm c}/v_{\rm imp})$, giving rise to a limiting COR $e_{\rm inf}$ at $v_{\rm imp}\rightarrow\infty$ and a critical impact velocity $v_{\rm c}$ below which no rebound would occur. The decrease of the COR with the increase of $h_{\rm liq}$ clearly suggests the influence of the viscous damping, since the energy dissipated through a viscous force depends on the distance that the particle travels. Different from the case of glass beads \cite{Gollwitzer12}, the COR for the first rebound is smaller than the fitted line for all film thicknesses. This could be attributed to the much stronger influence from the mass of liquid added to the particle during the first bouncing for the less dense PE sphere.

\begin{figure}
 \label{e_st}
 \includegraphics[height=.26\textheight]{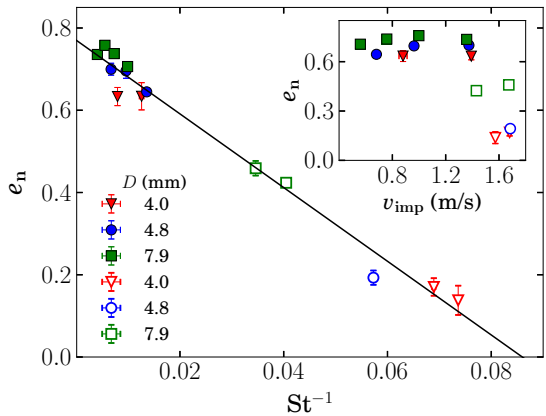}
 \caption{Normal coefficient of restitution $e_{\rm n}$ against the inverse Stokes number $\rm St^{-1}$ for a fixed $\tilde{h}\approx0.04$. Data with closed and open symbols present measurements with silicone oil M5 and M50 correspondingly. The inset shows the same data, but plotted against $v_{\rm imp}$.}
\end{figure}

Besides the liquid film thickness, the COR depends also strongly on the viscosity of the wetting liquid. As shown in the inset of Fig.~\ref{e_st}, the COR measured with silicone oil M50 as wetting liquid yields much lower values than those measured with M5. A re-plot of the COR with ${\rm St}^{-1}$ in Fig.~\ref{e_st} denotes a linear decay, suggesting that the dependence of the COR on $h_{\rm liq}$, viscosity $\eta$ and the impact velocity $v_{\rm imp}$ could be characterized by the Stokes number for fixed $\tilde{h}$. A linear fit with $e_{\rm n}=e_{\rm inf}(1-{\rm St_{c}}/{\rm St})$ yields a saturated COR $e_{\rm inf}=0.77\pm 0.04$ and a critical Stokes number $\rm St_{c}=11.64\pm 1.55$. Following a former theory \cite{Gollwitzer12}, we could estimate $e_{\rm inf}=e_{\rm dry}-3\tilde{\rho}\tilde{h}/2e_{\rm dry}=0.81\pm 0.03 $, which agrees with the measurement within the experimental uncertainties. This agreement suggests that the fact that $e_{\rm inf}$ is smaller than $e_{\rm dry}$ is due to the inertia of the wetting liquid, which provides the dependence on $\tilde{\rho}$ and $\tilde{h}$.  

To conclude, the normal coefficient of restitution for the impact between a spherical particle and a flat surface covered with a liquid film is investigated for various impact velocities, particle and liquid properties. The technique to determine the film thickness optically and the calibration results are described in detail. Compared with the former methods \cite{Gollwitzer12}, it provides an extended upper limit of the film thickness, and also allows the possibility to estimate the volume of the meniscus. At a certain density and size ratio, the COR is found to decay linearly with $\rm St^{\rm -1}$, which represents the scaling with the impact velocity and various particle as well as liquid properties. The parameter $e_{\rm inf}$ obtained from the fitting, corresponding to the saturated value of the COR, can be understood by a model considering the inertia of the liquid. A more detailed comparison to the model with various combinations of $\tilde{h}$ and $\tilde{\rho}$ will be a focus of our further investigations.


We thank Laura Meissner for performing the dry COR measurements. We are grateful for the support from Deutsche Forschungsgemeinschaft through HU1939/2-1.



\bibliographystyle{aipproc}   


\end{document}